\documentclass[preprint,12pt]{elsarticle}

\usepackage{geometry}
\usepackage{amssymb}
\usepackage{amsmath}
\usepackage{empheq}
\usepackage{tikz}
\usetikzlibrary{patterns,calc}
\usepackage{subcaption}

\usepackage{tabularray}
\usepackage{xcolor}
\usepackage{multirow}
\usepackage{stmaryrd}
\usepackage{colortbl}
\usepackage{empheq}

\usepackage{placeins}

\usepackage{mathrsfs}
\usepackage{graphicx}
\usepackage{bm}

\newcommand{\norm}[1]{\big|\big|#1 \big|\big|}
\newcommand{\dtch}[1]{\dfrac{\partial #1}{\partial t}}

\newcommand{\bch}[1]{\boldsymbol{#1}}
\newcommand{\udu}[2]{#1 \cdot \nabla #2}
\newcommand{\bu}{\boldsymbol{u}}
\newcommand{\mt}[1]{\mathcal{#1}}

\newcommand{\bfx}{\mathbf{x}}

\newcommand{\bfX}{\mathbf{X}}

\journal{arxiv}

\begin{document}

\begin{frontmatter}



\title{Identification of Settling Velocity with Physics Informed Neural Networks For Sediment Laden Flows}

\author[lemta,cor1]{Mickaël Delcey}
\ead{mickael.delcey@univ-lorraine.fr}
\author[lemta]{Yoann Cheny}
\cortext[cor1]{Corresponding author}
\author[JK]{Jean-Baptiste Keck}
\author[lemta]{Adrien Gans}
\author[lemta,iuf]{Sébastien Kiesgen de Richter}

\affiliation[lemta]{organization={LEMTA, CNRS, Université de Lorraine},
            addressline={2, Avenue de la Forêt de Haye},
            city={Vandœuvre-lès-Nancy, B.P. 160},
            postcode={54504},
            country={France}}
\affiliation[iuf]{organization={Institut Universitaire de France (IUF)},
            city={Paris},
            country={France}}

\affiliation[JK]{organization={Laboratoire Jean Kuntzmann},city={
Saint-Martin-d'Hères
},
country={France}}



\begin{abstract}
Physics-Informed Neural Networks (PINNs) have shown great potential in the context of fluid dynamics simulations, particularly in reconstructing flow fields and identifying key parameters. In this study, we explore the application of PINNs to recover the dimensionless settling velocity for sedimentation flow. The flow involves sediment-laden fresh water overlying salt water, which is described by Navier-Stokes equations coupled with sediment concentration and salinity transport equations. Two cases are investigated: one where the training data contains the salinity and sediment concentration fields, and another where it contains the velocity field. For both cases, we investigate several flow regimes and show that the model is capable of inferring the unknown parameter and reconstructing the hydrodynamic field of the flow. The quality of the model inference is assessed by comparing it with numerical simulations from a high-fidelity semi-Lagrangian solver. We demonstrate the model's robustness to noise by training it with data corrupted by noise of varying magnitudes, highlighting the potential of PINNs for real-world applications.

\end{abstract}




\end{frontmatter}


\section*{Introduction}\label{intro}

The settling of granular particles in a fluid is a widespread natural phenomenon with significant industrial implications. For instance, in mining, it aids in mineral extraction by segregating particles according to density, thereby improving efficiency. Pharmaceutical and food processing industries employ sedimentation in centrifugation processes to effectively separate components and refine liquids for quality assurance. In civil engineering, sedimentation principles in soil mechanics help comprehend groundwater flow dynamics, crucial for construction projects. Additionally, sedimentation plays a vital role in environmental engineering, notably in wastewater treatment, contributing to cleaner ecosystems and sustainable practices. Rivers serve as major conduits for transporting sediment from land to coastal oceans \cite{milliman1992geomorphic}, carrying pollutants such as microplastics \cite{goral2023settling}. Understanding the fundamental transport properties of these sediments is critical for addressing environmental concerns. Particularly, in situations where a lighter fluid laden with particles (freshwater) meets a denser clear fluid (saltwater), such as in estuarine environments, instabilities like Rayleigh-Taylor instability and double diffusivity emerge \cite{tory1996sedimentation}, leading to the formation of sedimentation fingers. The resulting downward sediment flux can be orders of magnitude above the one that would normally be obtained with only gravitational settling \cite{hoyal1999influence,parsons2000enhanced}. The critical factor influencing these instabilities has been identified as the ratio between the settling velocity of particles and the diffusive spreading velocity of the initial stable stratification layer \cite{burns2015sediment}. While the settling velocity of particles can be estimated based on their physical properties, the diffusivity of salinity ($\kappa_s$) relies entirely on the Schmidt number ($S_c$). The upward spreading velocity of the layer is influenced not only by $S_c$ but also by other dimensionless parameters of the system, such as the stability ratio ($R_s$) and the diffusivity ratio ($\tau$) between salinity and sediments.

The settling velocity is a key parameter which is commonly required in sediment transport models.
However, determining this parameter accurately is still a major challenge \cite{goral2023settling}. Furthermore, it has been shown that in estuaries, many factors such as sediment concentration, salinity and turbulent mixing can drastically affect the effective settling velocity \cite{manning2013factors}.

Physics  informed neural networks (PINNs) \cite{raissi2019physics}, although recent, has been established as a powerful tool to solve forward and inverse problems of partial differential equations (PDE) \cite{karniadakis2021physics}. In a PINN model, a neural network is used to approximate the solution of a PDE system. To this end,  the residuals of the governing equations computed by automatic differentiation and embedded in the loss function of the network as a regularizing term. As a consequence, by minimizing the loss, the network is constrained to satisfy the physical conservation laws. PINNs are an ideal solution for tackling inverse problems since it can handle complexe ill-posed problems where training data is incomplete and noisy, while also working well in complex geometries. PINNs has been widely used to tackle such inverses problems in computational science and engineering (e.g. fluid mechanics \cite{cai2021physics}, heat transfert \cite{cai2021physics}, power systems \cite{huang2022applications}).
Among the inverse problems, the identification of unknown parameters involves inferring not only the components of the solution vector $\bch{u}$ and their derivatives, but also constant parameters in the PDEs (e.g. coefficient such as viscosity and diffusivity), for example.
In \cite{cai2021heattransfert}, add two extra weights to PINN model to infer unknow constant in the boundary condition for stephan problem. In \cite{yin2021non}, the permeability and viscoelastic modulus are identified on the Cahn-Hilliard and Navier-Stokes equations, highlighting the possibility to infer phisiological parameters with PINNs.

In this work, we employ physics informed neural networks models to infer the dimenssionnless settling velocity in the context of particle-laden flows, governed by Navier-Stokes coupled with sediment concentration and salinity transport equations. This is to our knowledge the first attempt to identify settling velocity using the paradigm of PINNs. We investigate several flow regimes with varying settling velocity and Schmidt number, with different sets of training data as well as the influence of noise on the results accuracy. 
In the section 1, we describe the governing equations of the flow and the construction of references datasets for for several flow regimes involving various settling velocity $V_p$ and Schmidt number $Sc$. In the following section, we explain the principles of physics informed neural networks for flow reconstruction and its extention to parameter identification for double diffusive sedimentation. The section three introduce the results of the inference for each flow regimes.
The final section focuses on one specific flow regime to dive further in the study of results.

 \section{Problem setup}

 \subsection{Governing equations}
\begin{figure}[!ht]
    \centering
    \includegraphics[scale= 0.30
    ]{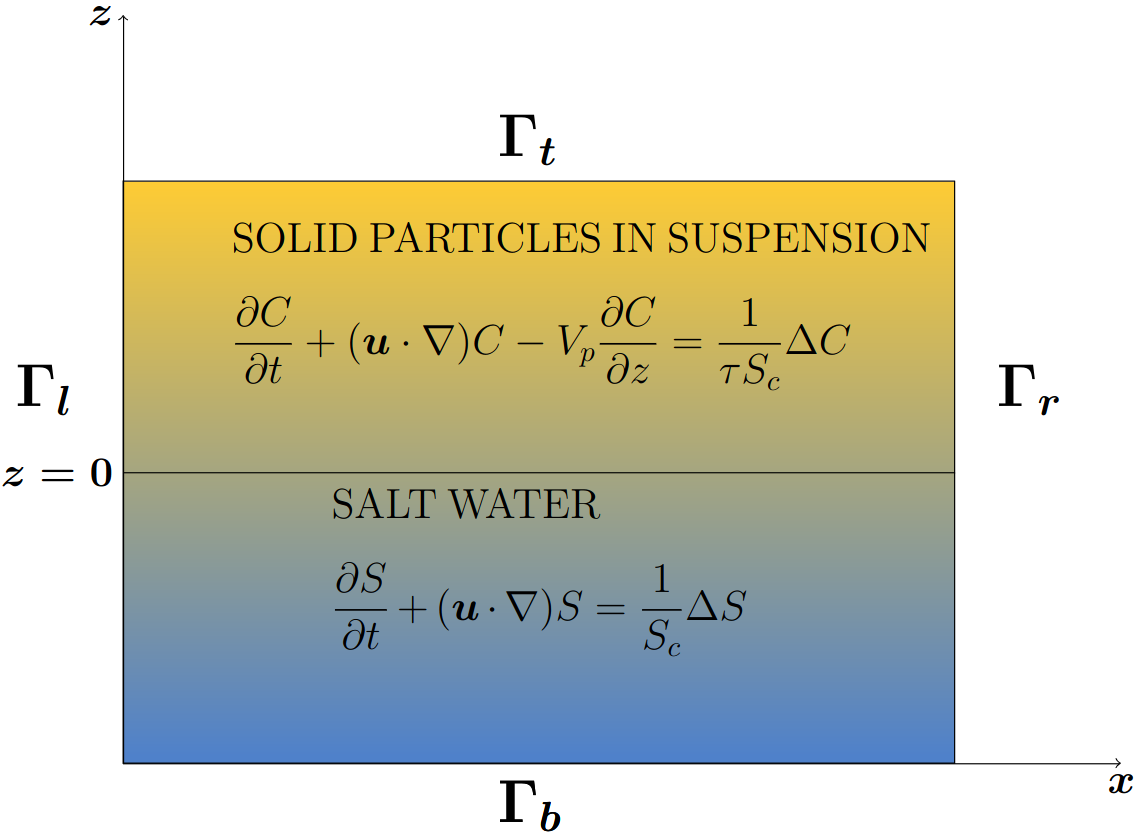}
    \caption{Two-Dimensionnal problem configuration.  $\boldsymbol{\Gamma}$ refers to the walls of the domain.}
    \label{fig:sketch}
\end{figure}

 We consider the same problem set-up as discussed in \cite{keck2021double} and \cite{burns2015sediment}. In this configuration, a layer of particle-laden fresh water is placed above clear, saline water. The horizontal directions are denoted $(x,y)$ in 3D and $x$ in 2D, while the vertical coordinate is represented by z. Initially, the interface between the two layers is situated at z = 0. The problem set-up is illustrated Fig (\ref{fig:sketch}). 

 The overall density $\rho$ is assumed to have a linear dependence on both salinity S and sediment concentration C, characterized by density expansion coefficients $\alpha$ and $\gamma$, respectively.  
\begin{align}
    \rho  = \rho_0(1 + \alpha S + \gamma C).
\end{align}

The sediment consists of small monodisperse particles with negligible inertia, which are carried by the combination of the fluid velocity $\bch{u}$ and the dimensionless settling velocity~$V_p$.

Under the Boussinesq approximation and within the framework of the velocity-vorticity formulation, the dimebsionless system is governed by the following dimensionless governing equations :

\begin{subequations} \label{eq:sediment_2}
    \begin{empheq}{gather} 
        \dfrac{\partial \bch{\omega}}{\partial t} + (\bch{u} \cdot \nabla ) \bch{\omega} = \Delta \bch{\omega} + (\bch{\omega} \cdot \nabla ) \bch{u} - \nabla \times (R_s + C)\bch{e_z},  \\
        \dfrac{\partial S}{\partial t} + (\bch{u} \cdot \nabla ) S = \dfrac{1}{S_c}\Delta S, \\\label{eq:sedoment_2:Vp}
        \dfrac{\partial C}{\partial t} + (\bch{u} \cdot \nabla ) C -V_p \dfrac{\partial C}{\partial z} = \dfrac{1}{\tau S_c} \Delta C, \\ \label{eq:sedoment_2:vort}
        \omega = \nabla \times \bch{u}, \\
        \nabla \cdot \bch{u} = 0,
        \end{empheq}
\end{subequations}
where $\bch{w} = (\omega_x, \omega_y, \omega_z)$ denotes the vorticity field. In 2D, the vorticity field 2D the vorticity field is reduced to a scalar field. The dimensionless parameters $R_s,~S_c,~ V_p$ and $\tau$ are introduced in table (\ref{tab:coefadim}). 

 \begin{center}
\begin{table}[h]
    \centering
    \begin{tabular}{cccc} 
    \hline
    Stability ratio & Diffusivity ratio & Schmidt number  & Settling velocity\\
     \hline 
    $R_s$ = $\dfrac{\alpha S_{max}}{\gamma C_{max}}$ &  
    $\tau$ = $\dfrac{\kappa_s}{\kappa_c}$ &  
    $S_c = \dfrac{\nu_0}{\kappa_c}$ & $V_p$ \\
    \hline
    \end{tabular}
    \caption{Dimensionless parameters. Here  $\alpha S_{\textrm{max}},~\gamma C_{\textrm{max}}$ refer to the maximum added density due to salinity and particles, $\kappa_s$, $\kappa_c$ are the diffusivity coefficients of salinity and sediments, and $\nu_0$ is the kinematic viscosity of fresh water.}
    \label{tab:coefadim}
\end{table}
\end{center}

\subsection{Numerical Setup}

The boundary conditions in the horizontal directions, denoted by $\Gamma_l$ and $\Gamma_r$ in fig. (\ref{fig:sketch}) are periodic. Top and bottom walls denoted $\Gamma_t$ and $\Gamma_b$ respectively are slip walls with no penetration. Furthermore, we impose no-flux boundary condition on those walls for the salinity field. For the sediment concentration, no-fluw boundary condition with respect to the settling velocity on $\Gamma_t$ and no-flux boundary condition on  $\Gamma_b$. 

\begin{subequations} \label{eq:bnd}
    \begin{empheq}{gather}
        \dfrac{\partial \bch{u}_{||}}{\partial z} = 0~ \textrm{and}~u_z = 0~ \textrm{on}~ \Gamma_t \cup \Gamma_b, \\
        V_p C - \dfrac{1}{\tau S_c} \dfrac{\partial C}{\partial z} = 0~\textrm{on}~\Gamma_t~\textrm{and}~ \dfrac{\partial C}{\partial z} = 0~\textrm{on}~\Gamma_b, \\
        \dfrac{\partial S}{\partial z} = 0~ \textrm{on}~\Gamma_t \cup \Gamma_b.
    \end{empheq}
\end{subequations}
Here, $\bch{u}_{||}$ represents the horizontal components $(u_x, u_y)$ in 3D and $u_x$ in 2D.

The initial condition field is given by a smoothed step profile. The initial salinity field is obtained by computing : $S = 1-C$.

\begin{subequations} \label{eq:init}
\begin{empheq}{gather} 
\bch{u}_0(x,z) = \bch{0}~\textrm{and}~\omega_0(x,z) = 0, \\
C_0(x,z) = \dfrac{1}{2} \left[ 1+ erf\left( \dfrac{z-\delta(x)}{l_0}\right) \right], \\
S_0(x,z) = 1 - C_0(x,z).
\end{empheq}

Here $\delta$ is a random perturbation uniformly distributed. The parameter $l_0$ is set to 1.5 for this simulation. As depicted in Fig. (\ref{fig:verical_profile}) the initial sediment concentration and salinity denoted $C_0$ and $S_0$, are shown with respect to the vertical coordinate $z$. 

\begin{figure}[h!]
    \centering
    \includegraphics[scale = 0.6]{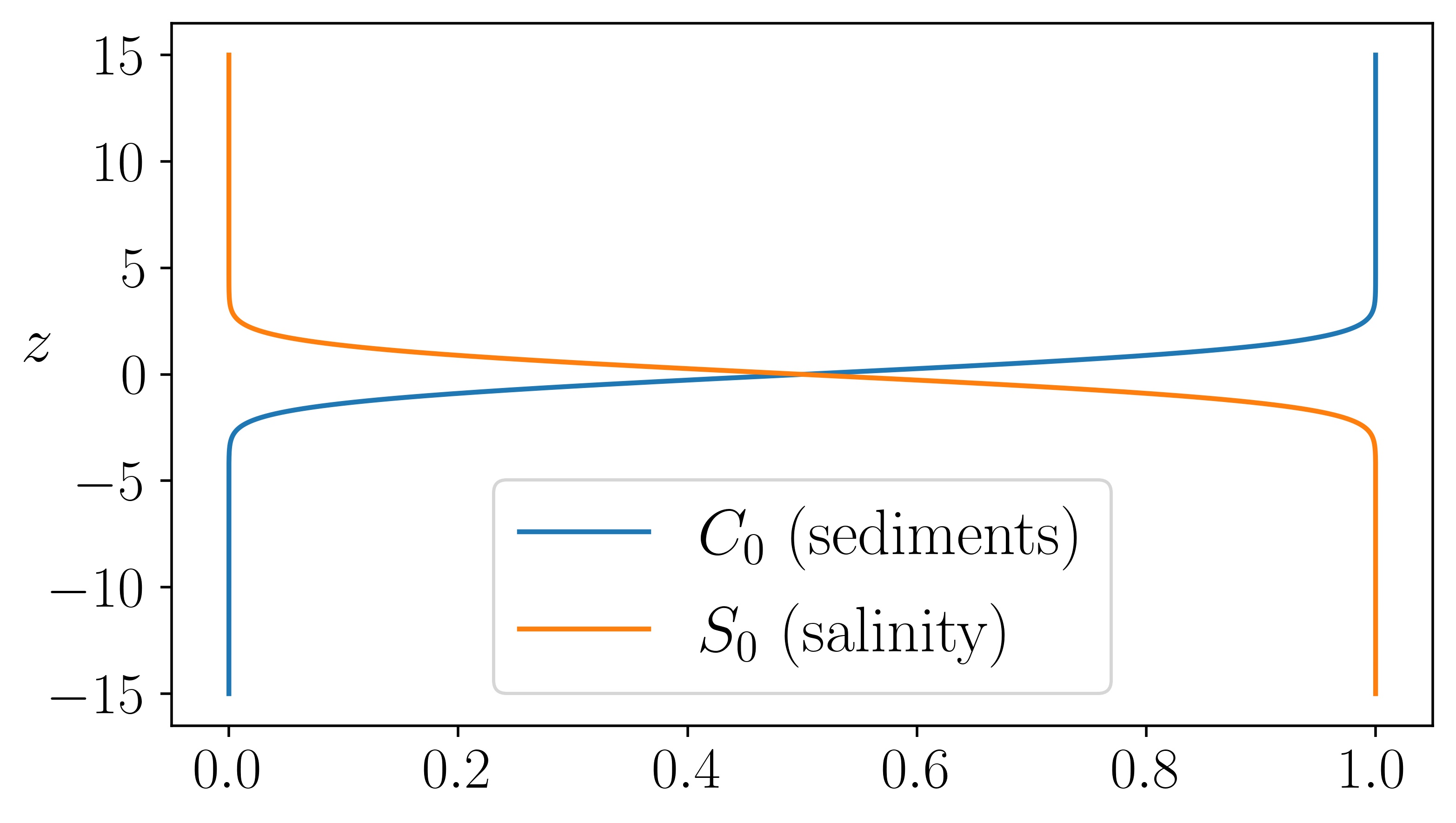}
    \caption{Initial vertical sediment profile for $l_0=1.5$, $\delta = 0.$ . }
    \label{fig:verical_profile}
\end{figure}
\end{subequations}

\begin{figure}[h]
    \centering
    \includegraphics[scale = 0.50]{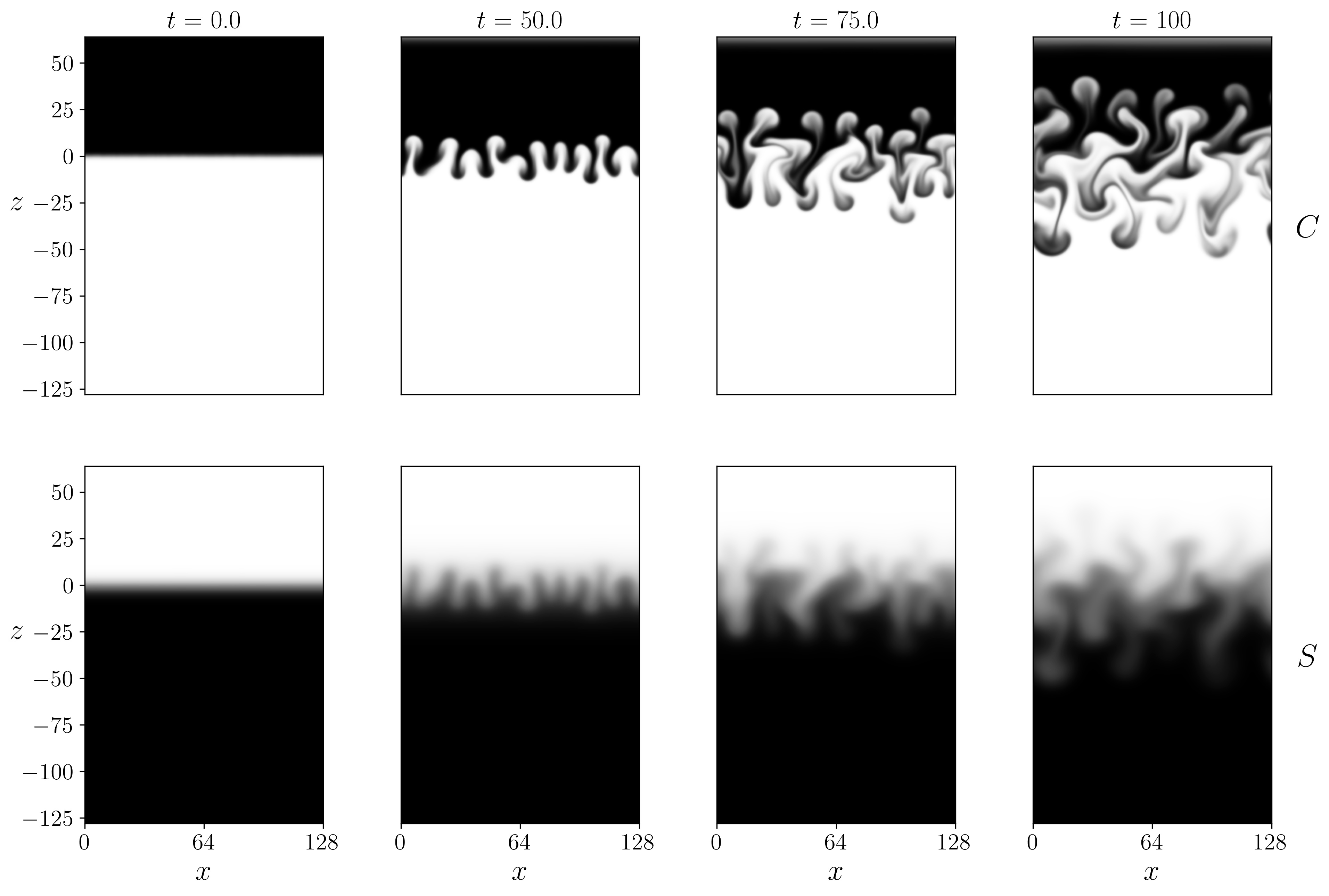}
    \caption{Snapshot of the sediment concentration (first line) and  salinity (second line) at dimensionless times $t=0,~50,~75,~100~$. Parameters : $V_p = 0.04$, $S_c = 0.7$, $R_s = 2$, $\tau = 25$.}
    \label{fig:visualisation}
\end{figure}

We solve the 2D version of equations (\ref{eq:sediment_2}) with boundary conditions (\ref{eq:bnd}) and initial condition (\ref{eq:init}) using the highly accurate semi-lagrangian method described in \cite{keck2021double}. The numerical solution is obtained on the dimensionless computational domain  $\Omega = [0,128] \times [-128, 64]$ for $t \in [0,100]$. The evolution of sediment concentration and salinity fields is illustrated Fig. (\ref{fig:visualisation}).

Several numerical simulations are performed using fixed parameters $Rs=2$ and $\tau=25$, while varying the Schmidt number and settling velocities as detailed in Table (\ref{tab:regimes}).

\begin{table}[h]
    \centering
    \begin{tabular}{c|cccccccc}
    \hline
         $S_c$ & 0.7 & 3.5 & 7.0 & 7.0 & 7.0 & 7.0 & 14.0 & 28.0\\
         $V_p$ & 0.04 & 0.04 & 0.01 & 0.04 & 0.08 & 0.16 & 0.04 & 0.04 \\
    \hline
    \end{tabular}
    \caption{Flow configuration }
    \label{tab:regimes}
\end{table}

\clearpage

\subsection{Syntethic datasets}

For each case, we construct a reference dataset within the same spatio-temporal framework, which is utilized for training PINNs. We select a spatial subdomain $\Omega_1 \subset \Omega$, and the subset $\Omega_1 \times [0,T]$ is discretized into a uniform Cartesian grid consisting of $N_x \times N_z \times N_t = 342 \times 128 \times 125$ points. We denote the set containing all the spatio-temporal coordinates of the discretized grid as $\mathcal{C}= {\mathbf{x}_i, t_i }$, and $\mathcal{D} = {S_i, C_i, \mathbf{u}_i, \omega_i }$ represents the set containing the corresponding solution fields for given $S_c$ and $V_p$ parameters.

As we plan to conduct multiple PINNs trainings across all flow regimes outlined in Table \ref{tab:regimes}, the dataset $\mt{D}$ is intentionally kept smaller than the numerical simulation dataset to reduce computational costs. Furthermore, we have trimmed the spatial domain to exclude non-essential values from the training process.

\begin{figure}[h]
    \centering
    \includegraphics[scale = 0.5]{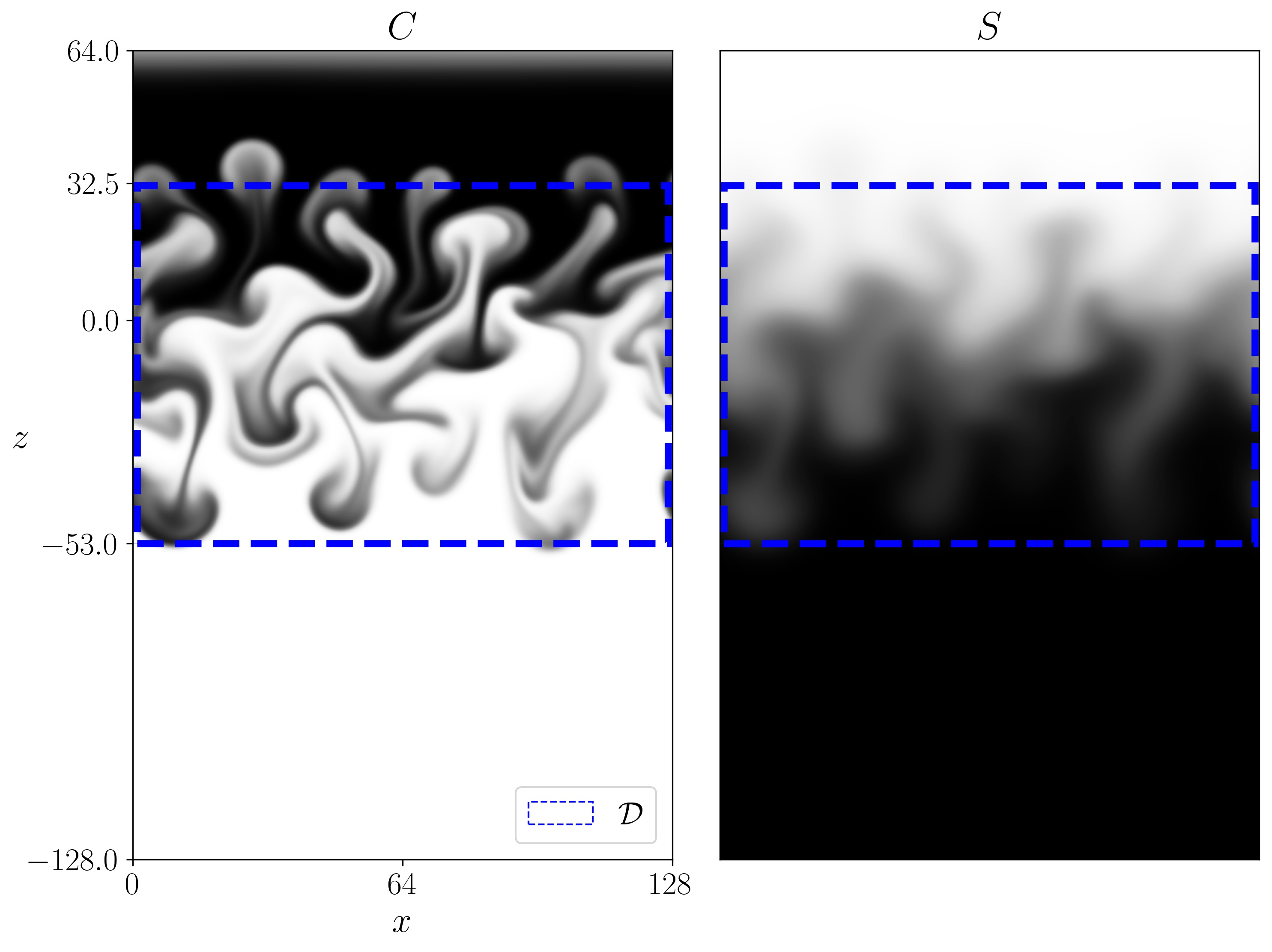}
    \caption{}
    \label{fig:datasets}
\end{figure}

\section{Physics Informed Neural Networks}

\subsection{Inverse problem}

In this section, we delve into the principles of Physics-Informed Neural Networks (PINNs) within the context of sedimentation flow. Our objective with the PINN model is to deduce missing information about the flow dynamics by leveraging the governing equations \eqref{eq:sediment_2} along with available observations of the solution field, specifically $(S, C, \mathbf{u}, \mathbf{\omega})$.

To this end, an Artificial Neural Network (ANN) denoted $\mt{N}$, is used to infer the solution of (\ref{eq:sediment_2}) while approximating a set of observed data. \\The network is a function that take as inputs $(\bch{x},t) \in \Omega \times [0,T]$ and predicts :
\begin{align}\label{Net}
    \mt{N}(\bch{x},t;\theta) = (S_{\mt{N}},C_{\mt{N}},\bch{u}_{\mt{N}},\bch{\omega}_{\mt{N}})(\bch{x},t;\theta),
\end{align}
the parameter $\theta$ contains the trainable weigths and biases of the network $ \mt{N}$.\newline The loss function of the ANN contains two terms : The first is the contribution of the data, denoted $\mt{L}_{\textrm{data}}$ which penalizes the mismatch between the measured data and the corresponding network predictions. The second, denoted $\mt{L}_{\bch{f}}$ is associated to the conservations laws.

We define  :

\begin{empheq}[left = \bch{f}_{\mt{N}}(\bch{x}\textrm{,}t\textrm{;}\theta)\textrm{=}\empheqlbrace]{align*}
              &\displaystyle \dtch{\bch{\omega}_{\mt{N}}} + ( \bch{u}_{\mt{N}} \nabla) \bch{\omega}_{\mt{N}} 
              - (\bch{\omega}_{\mt{N}} \cdot \nabla )\bch{u}_{\mt{N}} = \Delta \bch{\omega}_{\mt{N}} - \nabla \times ( R_s S_{\mt{N}} + C_{\mt{N}} ) \bch{e_z}\\
              &\dtch{S_{\mt{N}}} + (\bch{u} \cdot \nabla )S = \frac{1}{S_c}\Delta S \\
              &\dtch{C_{\mt{N}}} + (\bu_{\mt{N}}\cdot \nabla)C_{\mt{N}} - V_p\frac{\partial C_{\mt{N}}}{\partial z} = \frac{1}{\tau S_c} \Delta C_{\mt{N}} \\
              &\displaystyle \bch{\omega}_{\mt{N}} = \nabla \times \bch{u}_{\mt{N}} \\
            &  \nabla \cdot \bch{u}_{\mt{N}} 
\end{empheq}

The function $\bch{f}_{\mt{N}}$ is the residuals of equations (\ref{eq:sediment_2}) given by an approximation of equations (\ref{eq:sediment_2}) by $\mt{N}$, the derivatives involved in $\bch{f}_{\mt{N}}$ are computed by Automatic Differentiation \cite{baydin2018automatic}, which allows pointwises derivatives without tuncation errors. 
The loss function is expressed as :

\begin{empheq}{gather}\label{eq:lossfct}
    \mt{L}(\theta) = \lambda\mt{L}_{\textrm{data}}(\theta) + \mt{L}_{\bch{f}}(\theta)
    ,
\end{empheq}

where
\begin{empheq}{align}\label{eq:residuals}
    \mt{L}_{f}(\theta) = \dfrac{1}{N_{\bch{f}}}\displaystyle \sum_{i=1}^{N_{\bch{f}}} \norm{\bch{f}(\bch{x}^f_i,t^f_i;\theta)}_2^2 .
\end{empheq}
The parameter $\lambda$ is a constant set manually to balance the two error terms.

The training phase consists in finding the parameters of $\theta^*$ which minimizes the loss $\mt{L}$.
\begin{align}
    \theta^* = \underset{\theta}{argmin ~ \mt{L}(\theta)}.
\end{align}

In practice, the minimization process consists in iterating the gradient descent algorithm :
\begin{empheq}{align}
    \theta_{n+1} = \theta_{n} - \eta \nabla_{\theta} \mt{L},
\end{empheq}
where $\eta$ is the learning rate. \\
In the field of machine learning, the stochastic gradient descent algorithm \cite{bottou2010large} is commonly chosen to mitigate computational costs. This method involves updating the model's parameters iteratively, using batches of data. Each batch provides an estimation of the overall gradient based on a subset of the dataset. This contributes to faster convergence and more efficient use of computational resources compared to the traditional gradient descent approach, where the entire dataset is used for each update. The term used to describe the process of completing one pass through the entire dataset is commonly referred to as an \emph{epoch}.

\subsection{Parameter Identification}

Parameter identification is a special case of inverse problems for PINNs, where missing data are not only components of the hydrodynamic field, but could also be constant parameters in (\ref{eq:sediment_2}) such as $R_s,~ B_p,~ S_c$ or $\tau$. 
In our case, the objective is to deduce the dimensionless settling velocity $V_p$ from partial data on the flow fields.

Thus, we assume that $V_p$ is now an unknown variable. In the PINN framework, we introduce a trainable parameter into the network, denoted $\tilde{V_{p}}$ so that the function $\bch{f}_{\mt{N}}$ is now expressed as :

\begin{empheq}[left = \bch{f}_{\mt{N}}(\bch{x} \textrm{,}~t\textrm{;}~\Theta)\textrm{=}\empheqlbrace]{align*}
              &\displaystyle \dtch{\bch{\omega}_{\mt{N}}} + ( \bch{u}_{\mt{N}} \nabla) \bch{\omega}_{\mt{N}} - (\bch{\omega}_{\mt{N}} \cdot \nabla )\bch{u}_{\mt{N}} = \Delta \bch{\omega}_{\mt{N}} - \nabla \times ( R_s S_{\mt{N}} + C_{\mt{N}} ) \bch{e_z}\\
              &\dtch{S_{\mt{N}}} + (\bch{u} \cdot \nabla )S = \frac{1}{S_c}\Delta S \\
              &\dtch{C_{\mt{N}}} + \udu{\bu_{\mt{N}}}{C_{\mt{N}}} - \Tilde{V_{p}}\frac{\partial C_{\mt{N}}}{\partial z} = \frac{1}{\tau S_c} \Delta C_{\mt{N}} \\
            &\displaystyle \bch{\omega}_{\mt{N}} = \nabla \times \bch{u}_{\mt{N}} \\
            &  \nabla \cdot \bch{u}_{\mt{N}} 
\end{empheq}
Here, $\Theta = (\theta, \Tilde{V_{p}})$ contains the previous weights of the network $\mt{N}$ but also a new trainable weigth $\Tilde{V_{p}}$ used to approximate the real value of $V_p$ during training.

As a consequence, this parameter is involved in the second loss term $ \mt{L}_{f}$ and the global loss function of the network stay the same as (\ref{eq:lossfct}).

\begin{figure}[]
    \centering
    \includegraphics[scale = 0.285]{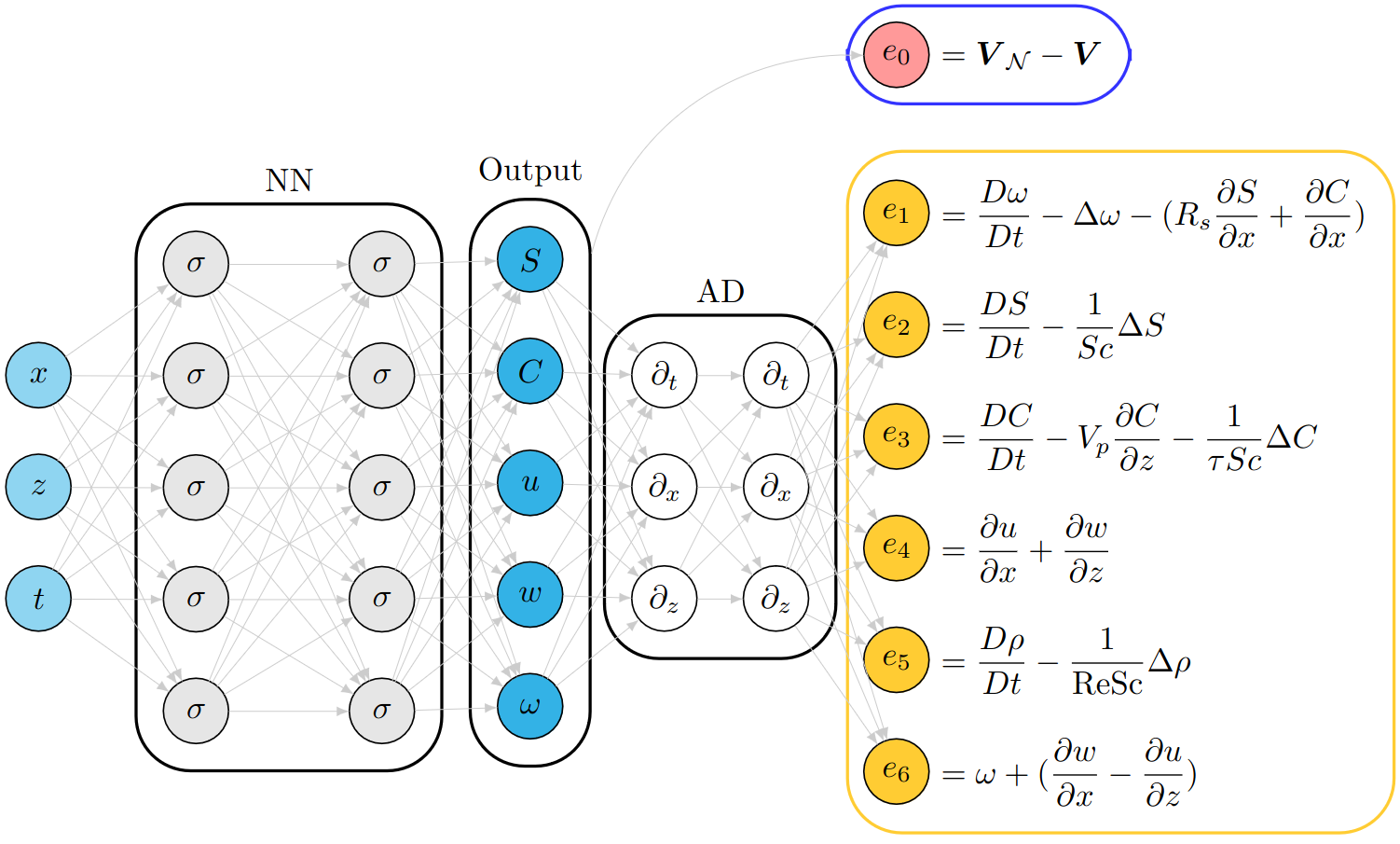}
    \caption{Physics-informed neural network structure : a fully connected neural network take as input $(\bch{x},t) \in \mathbb{R}^{3}$ and predicts $\boldsymbol{\mathcal{N}}(\bch{x},t) = (S_{\mathcal{N}},C_{\mathcal{N}},u_{\mathcal{N}},w_{\mathcal{N}},\omega_{\mathcal{N}})$. The residuals of the governing equations $e_{1,6}$  are computed by automatic differentiation and $e_0$ denotes the mismatch between the observational data $\bch{V}$ and the predictions $\bch{V}_{\mathcal{N}}$, which are combined in the loss function $\mathcal{L}$. Finally, the weights and biases of the network are updated iteratively with gradient descent.  }
    \label{fig:sketchPINN}
\end{figure}

We can see an illustration of the network $\mt{N}$ in figure (\ref{fig:sketchPINN}).

\FloatBarrier

\clearpage

\section{Identification of settling velocity accross flow regimes}\label{section:regimes}

In this section, we use a physics-informed neural network model to deduce the settling velocity across various flow regimes outlined in the table (\ref{tab:regimes}) using incomplete data. The training process involves two distinct configurations of the training datasets.

\begin{itemize}
    \item In the first configuration, we assume that the observation data available are salinity and sediment concentration fields. In practical situations, these fields might correspond to measurements obtained through the PILAT technique, as discussed in \cite{schneider2023investigation}. This case is refered as \textbf{PILAT case}. 
    \item In the second, we use the velocity field as observationnal data. In real-world scenarios, these fields may correspond to measurements obtained using the so called PIV\footnote{Particle Image Velocimetry} technique \cite{raffel2007particle}.
    This case is refered as \textbf{PIV case}.
\end{itemize}

Therefore, for each training configuration, the loss function is formulated as in \eqref{eq:lossfct} where the term $\mathcal{L}_{\textrm{data}}$ depends on the training data. For the PILAT case, $\mt{L}_{\textrm{data}}$ addresses the discrepancy between the observed salinity and sediment concentration and the corresponding fields predicted by the model :

\begin{empheq}{align}\label{eq:Loss_SC}
    \mathcal{L}_{\textrm{data}} &= \mt{L}_{S} + \mt{L}_C=\dfrac{1}{N}\sum_{i=1}^{N} \norm{S_{\mt{N}}(\boldsymbol{x}_i,t_i)- S_i}^2 +
    \dfrac{1}{N}\sum_{i=1}^{N} \norm{C_{\mt{N}}(\boldsymbol{x}_i,t_i) - C_i}^2,
\end{empheq}
while for the PIV case, $\mt{L}_{\textrm{data}}$ is the mismatch between observed and predicted velocity :

\begin{empheq}{align}\label{eq:Loss_UV}
    \mathcal{L}_{\textrm{data}} &= \mt{L}_{u} + \mt{L}_w=\dfrac{1}{N}\sum_{i=1}^{N} \norm{u_{\mt{N}}(\boldsymbol{x}_i,t_i)- u_i}^2 +
    \dfrac{1}{N}\sum_{i=1}^{N} \norm{w_{\mt{N}}(\boldsymbol{x}_i,t_i) - w_i}^2 .
\end{empheq}

Here $\forall i \in \llbracket 1, N\rrbracket,~ u_i,v_i,S_i,C_i \in \mt{D}$, we use the same set of points $(\bch{x}_i, t_i) \in \mt{C}$ to compute the residuals $\mt{L}_f$ and $\mt{L}_{\textrm{data}}$ for each cases. Furthermore, the parameter $\lambda$ is set to $1000$.

For each training, we use a fully connected ANN with 9 hidden layers of 195 neurons each. \\
The activation function \emph{swish} \cite{swishramachandran2017searching} is selected for all layers except the last one where no activation function is imposed. The loss function is optimized during training using Adam method \cite{kingma2014adam} with a batch size of 1024 for 70 epochs. The learning rate of the gradient descend decreases from $5 \times 10^{-4}$ to $1 \times 10^{-5}$. Finally, we create an additionnal trainable parameter $\Tilde{V_{p}}$ which refers to the inferred settling velocity. 

The results of the training are inverstigated using the relative $L_2$ error defined by :

\begin{equation}\label{eq:L2gene}
\epsilon_V =\frac{100}{\sup_{\bfx \in \mt{C}} \left| V_\bfx \right|}\sqrt{\frac{\sum_{\bfx\in\bfX}	\left|V_{\mathcal{N}}(\bfx) - V_\bfx\right|^{2}}{\left|\bfX\right|}}
\end{equation}
where $V \in (s,c,u,w,\omega)$ and $\bfX$  defines the domain where the error is computed. We choose $\sup_{\bfx \in \Omega_{obs} \times T_{obs}} \left| V_\bfx \right|$ as reference value to avoid the division near to zero issue reported in \cite{pof}. \newline

\begin{table}[h]
    \centering
    \begin{tabular}{ cc|cccccc } 
     \hline
     $S_c$ & $V_p$ & $\Tilde{V_p}$ & $  \epsilon_s$ & $\epsilon_c$ & $\epsilon_u$ & $\epsilon_w$ & $\epsilon_\omega$ \\
     \hline 
     3.5 & 0.04 & 0.04 & 0.08\% & 0.30\% & 0.91\% & 0.92\% & 1.10\% \\
     7 & 0.01 & 0.0098 &0.06\% & 0.21\%  & 0.79\%  & 0.57\%  & 1.64\%   \\
     7 & 0.04 & 0.039 & 0.07\% & 0.23\% & 0.98\% & 0.60\% & 1.25\% \\
     7 & 0.08 & 0.079 & 0.14\% & 0.58\% & 0.98\% & 0.97\% & 1.48\% \\
     7 & 0.16 & 0.158 & 0.36\% & 1.53\% & 1.12\% & 0.81\% & 1.48\% \\
     14 & 0.04 & 0.039 & 0.08\% & 0.26\% & 0.89\% & 0.50\% & 1.34\% \\
     28 & 0.04 & 0.039 & 0.08\% & 0.25\% & 0.90\% & 0.42\% & 1.10\% \\
     \hline
    
    \end{tabular}
    \caption{Results for the differents flow regimes. $\Tilde{V_p}$ is the predicted dimensionless settling velocity and $\epsilon_{\cdot}$ refers to the $L_2-$ relative error (\ref{eq:L2gene}) between reference and predicted fields for PILAT case. Error $\epsilon$ is computed on all the spatio-temporal domain $\Omega_{1} \times [0, T]$.}
    \label{tab:results1}
\end{table}

The outcomes presented in table (\ref{tab:results1}) demonstrate that the model consistently converges to the accurate settling velocity across various scenarios, while also precisely inferring flow characteristics. Given that salinity and sediment concentration fields were included in the training data, their associated errors were anticipated to be minimal, as observed (below 2\%). Remarkably, errors related to velocity and vorticity fields also remained under 2\%, underscoring the model's great accuracy in reconstructing these fields.

Notably, alterations in the Schmidt number, while maintaining a constant settling velocity, did not significantly impact the outcomes. Conversely, an increase in the settling velocity $V_p$ led to augmented reconstruction errors across all fields. This trend may be attributed to the accelerated sedimentation at higher $V_p$ values, which intensifies flow patterns and instabilities, thereby explaining the observed increase in network reconstruction error for elevated settling velocities.

To rigorously evaluate the robustness of our model under varying conditions of data quality within the parameter identification framework, we introduced controlled levels of noise into the training datasets across all examined flow regimes. This approach was designed to simulate real-world data inaccuracies and test the resilience of our model's predictive capabilities.

We augmented the training data with Gaussian noise characterized by a zero mean ($\mu = 0$) and a standard deviation ($\sigma$) proportionate to the maximum value of the training field. Specifically, we varied $\sigma$ to represent 0\%, 15\%, 25\%, and 35\% of the field's maximum value, denoted as $\mathcal{N}(0, \sigma^{2})$. This stratification allowed us to systematically analyze the impact of noise on the model's performance across a spectrum of data quality scenarios. Thus, for PILAT case,

\begin{empheq}{align}\label{eq:Loss_SCN}
    \mathcal{L}_{\textrm{data}} 
    =\dfrac{1}{N}\sum_{i=1}^{N} \norm{S_{\mt{N}}(\boldsymbol{x}_i,t_i)- \Tilde{S}_i}^2 +
    \dfrac{1}{N}\sum_{i=1}^{N} \norm{C_{\mt{N}}(\boldsymbol{x}_i,t_i) - \Tilde{C}_i}^2 ,
\end{empheq}
where $\Tilde{S} = S + \xi_s,~ \xi_s \sim \mt{N}(0,\sigma^2) $ and  $\Tilde{C} = C + \xi_c,~ \xi_c \sim \mt{N}(0,\sigma^2) $ where $S,C \in \mt{D}$ and $\sigma$ is the standard deviation of the gaussian law.

\begin{figure}[h]
    \centering
    \includegraphics[scale = 0.35]{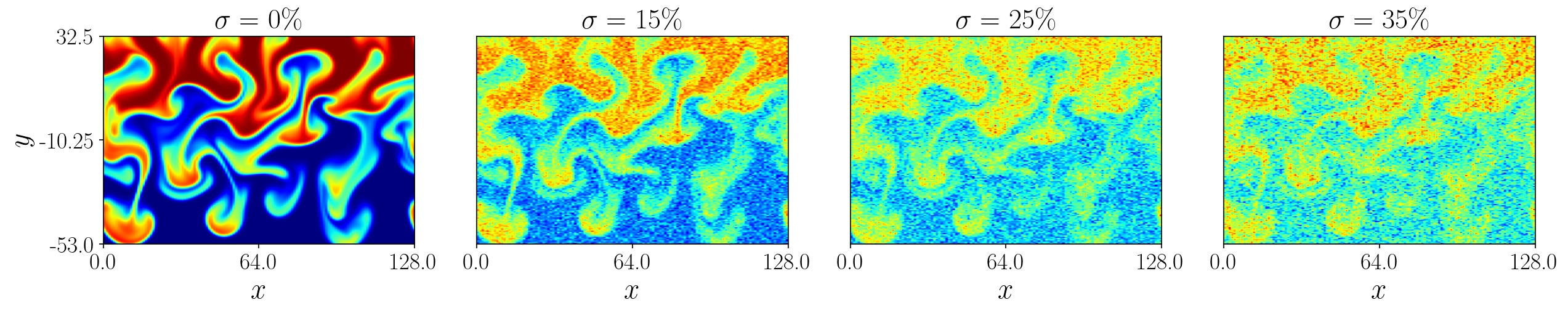}
    \caption{Visualisation of sediment concentration field data at $t=100$ corrupted by gaussian noise $\xi_{C}\sim \mt{N}(0, \sigma^2)$, where $\sigma =0\%,~15\%,~25\%,~35\%$}
    \label{fig:noise}
\end{figure}

For PIV case, 

\begin{empheq}{align}\label{eq:Loss_UVN}
    \mathcal{L}_{\textrm{data}} 
    =\dfrac{1}{N}\sum_{i=1}^{N} \norm{u_{\mt{N}}(\boldsymbol{x}_i,t_i)- \Tilde{u}_i}^2 +
    \dfrac{1}{N}\sum_{i=1}^{N} \norm{w_{\mt{N}}(\boldsymbol{x}_i,t_i) - \Tilde{w}_i}^2 ,
\end{empheq}
where $\Tilde{u} = u + \xi_u,~ \xi_u \sim \mt{N}(0,\sigma^2) $ and  $\Tilde{w} = w + \xi_w,~ \xi_w \sim \mt{N}(0,\sigma^2) $ where $u,w \in \mt{D}$.\\
A visualisation of the noisy sediment concentration field can be seen figure (\ref{fig:noise}).

The results are presented in Table (\ref{tab:noiseSC}) for PILAT and in (\ref{tab:noiseUV}) for PIV. Theses results showcase the robustness of PINNs when tackling the identification problem with noisy data. While for both cases it is generally observed that lower noise levels yield more accurate inferences, even under high levels of noise, the outcomes consistently align with the reference value of $V_p$.

However, as outlined in table (\ref{tab:noiseUV}), despite achieving close results to the PILAT cases in the absence of noise, the performance of the PIV case undergoes a significant drop when noise is introduced. This observation becomes apparent when examining the flow regime $Sc=0.7$, $V_p=0.16$. In this instance, the inferred value of Vp for the PILAT case is 0.1534, even with 35\% noise, while it drops to 0.14 for the PIV cases.

The decrease in performance may be attributed to the position of $V_p$ within the equations (\ref{eq:sediment_2}). Indeed, Equation (\ref{eq:sedoment_2:Vp}) is the only one that involves $V_p$, and in this equation, the sediment concentration field holds a significant role. Notably, the quantity $\frac{\partial C}{\partial z}$ is directly multiplied by Vp. Consequently, it might be more challenging for the network to infer the temporal and spatial derivatives of $C$ from noisy velocity data than for sediment concentration field itself.

\begin{table}[h]
    \centering
     \begin{tabular}{c |c |c| c| c | c}
    \hline 
    $S_c$  & $V_p$ & \multicolumn{4}{c|}{$\Tilde{V_{p}}$} \\
    \hline 
      \multicolumn{2}{c|}{} & $\sigma =0\%$ &  $\sigma = 15\%$ & $\sigma = 25\%$ & $\sigma = 35\%$\\
        \hline
        0.7 & 0.04 & 0.0401 & 0.0397 & 0.0399 & 0.0388 \\
        3.5 & 0.04 & 0.0402 & 0.04078 & 0.04011 & 0.0384\\
        7.0 & 0.01 & 0.00994 & 0.00957 & 0.0082 & 0.0077\\
        7.0 & 0.04 & 0.0401 & 0.0395 & 0.0388 & 0.0385\\
        7.0 & 0.08 & 0.0803 & 0.0799 & 0.0792 & 0.0759\\
        7.0 & 0.16 & 0.1604 & 0.1573 & 0.1568 & 0.1534\\
        14.0 & 0.04 &  0.0400 & 0.0393 & 0.0384 & 0.0389\\
        28.0 & 0.04 &  0.0397 & 0.0388 & 0.0389 & 0.0367\\
        \hline
    \end{tabular}
    \caption{Comparison between reference $V_p$ and  $\Tilde{V_p}$ predicted for all flow regimes of (\ref{tab:regimes}). Salinity et sediment concentrations $S$ and $C$ (PILAT cases) from reference dataset $\mt{D}$  are used as training data and corrupted by gaussian noise of varying magnitudes  $\mt{N}(0, \sigma)$ with $\sigma = 15\%,~ 25\% $ and $35\%$.}
    \label{tab:noiseSC}
\end{table}

\begin{table}[h]
    \centering
    \begin{tabular}{c |c |c| c| c | c}
    \hline 
    $S_c$  & $V_p$ & \multicolumn{4}{c}{$\Tilde{V_{p}}$}  \\
    \hline 
     \multicolumn{2}{c|}{} & $\sigma = 0\%$ & $\sigma = 15\%$ & $\sigma = 25\%$ & $\sigma = 35\%$\\
        \hline
        0.7 & 0.04 & 0.0395 & 0.0402 & 0.0375 & 0.0362 \\
        3.5 & 0.04 & 0.0397 & 0.0354 & 0.0351 & 0.0345 \\
        7.0 & 0.01 & 0.0097 & 0.0082 & 0.0089 & 0.0075  \\
        7.0 & 0.04 & 0.0390 & 0.0329 & 0.0295 & 0.0289 \\
        7.0 & 0.08 & 0.0789 & 0.0698 & 0.0682 & 0.0675 \\
        7.0 & 0.16 & 0.1593 & 0.133 & 0.132 & 0.140 \\
        14.0 & 0.04 & 0.0394 & 0.0338 & 0.0317 & 0.0295 \\
        28.0 & 0.04 & 0.0392 & 0.0313 & 0.0258 & 0.0312 \\
        \hline
    \end{tabular}
    \caption{Comparison between reference $V_p$ and  $\Tilde{V_p}$ predicted for all flow regimes of (\ref{tab:regimes}). Velocity components $u$ and $w$ (PIV case) are used as training data and corrupted by gaussian noise of varying magnitudes  $\mt{N}(0, \sigma)$ with $\sigma = 15\%,~ 25\% $ and $35\%$.}
    \label{tab:noiseUV}
\end{table}

\begin{figure}[h!]
    \centering
    \includegraphics[scale = 0.385]{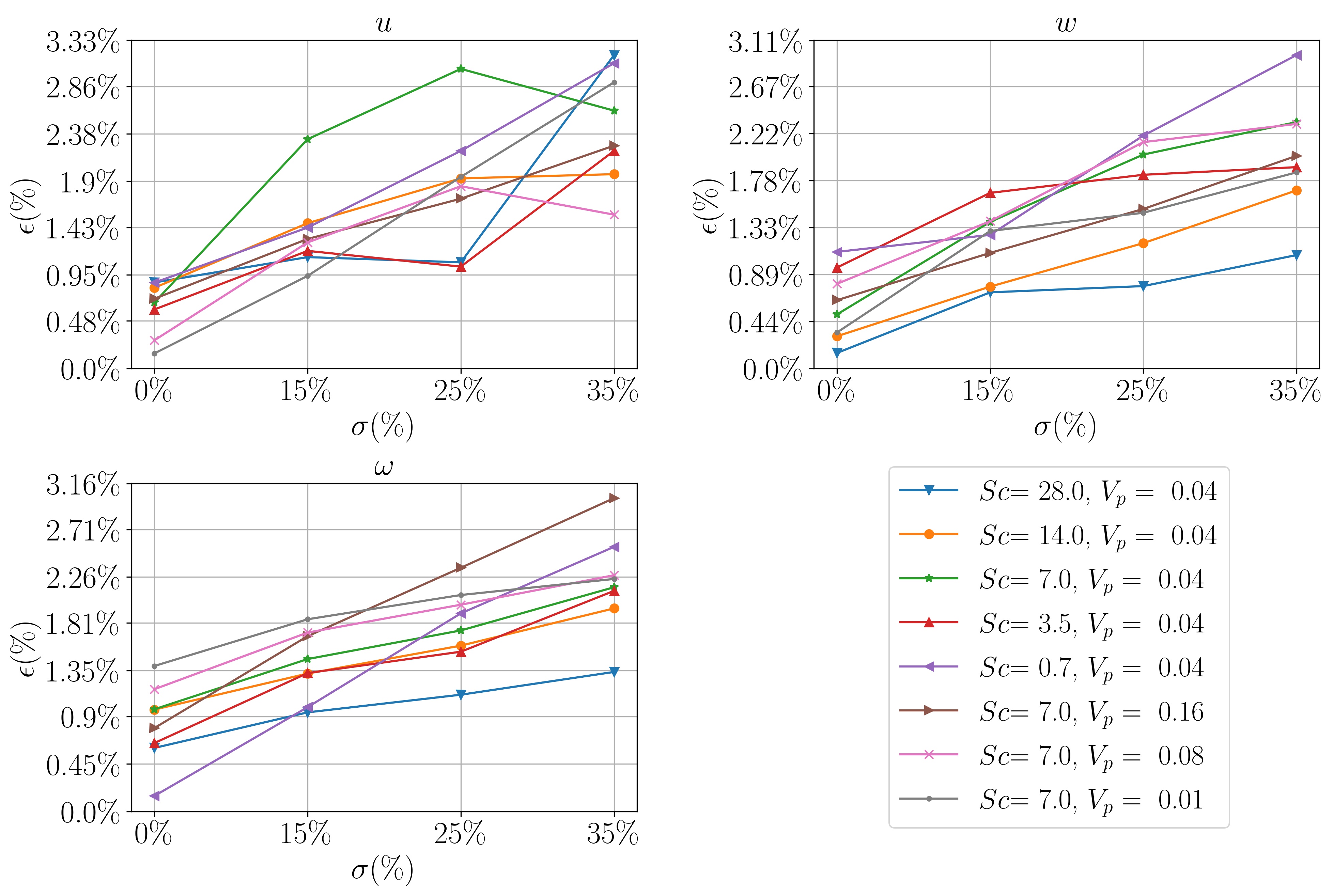}
    \caption{$L_2$-relative error (\ref{eq:L2gene}) predicted and reference velocity and vorticity. The error is computed on the discrete set of spatio-temporal coordinates $\mt{C}$ corresponding to the training domain $\Omega_{obs} \times [0, T_{obs}]$ of figure (\ref{fig:datasets}).  Each flow regimes of (\ref{tab:regimes}) are investigated. }
    \label{fig:error_noise}
\end{figure}

Additionally, in Figure (\ref{fig:error_noise}), we can observe the progression of the $L_2$-relative error for each studied flow regime in response to noise in the PILAT case. This error is specifically computed for velocity and vorticity fields that are not part of the training dataset.
We can see that in the majority of cases, higher levels of noise lead to deteriorating results, which is consistent with the conclusion regarding the influence of noise on $V_p$ identification. \newline 
\clearpage
 A more precise study of the identification process and reconstruction results will be undertaken in the next section.

\clearpage

\section{Case study : $V_p=0.04$, $Sc=0.7$. }

In this section, we delve further into the analysis of the identification problem, we focus on the scenario where $Sc$ equals 0.7 and $V_p$ is set to 0.04, which serves as a reference for the canonical scenario discussed in \cite{burns2015sediment}.
We keep the same network parameters as in section (\ref{section:regimes}) and the loss function (\ref{eq:Loss_SC}) which is computed on the cartesian grid $\mt{C}$.

\begin{figure}[h]
    \centering
    \includegraphics[scale = 0.4]{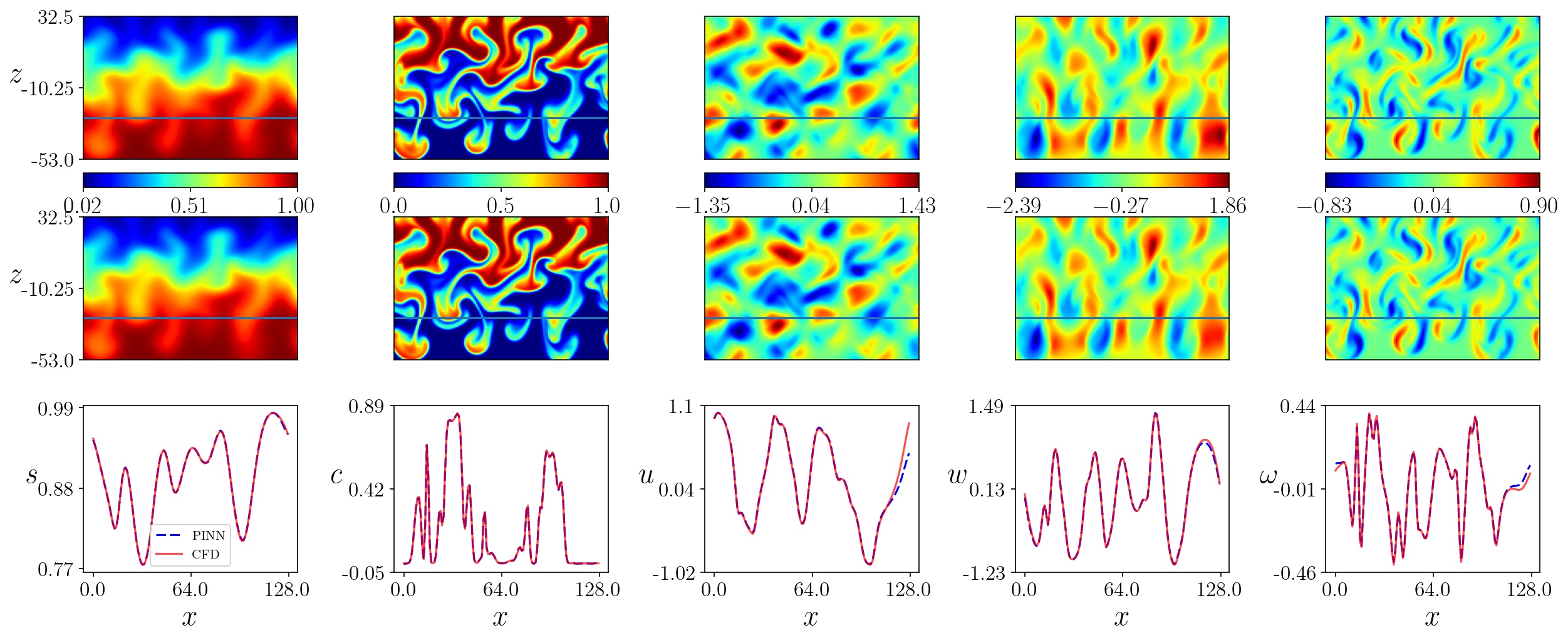}
    \caption{Comparison of horizontal profiles.  Snapshots of salinity, sediment concentration velocity and vorticity at $t = 100$ for the reference solution (first row) and the PINN prediction (seconde row). Horizontal profiles at $z = -28.0$ (red blue line on the snapshots) are compared between CFD reference solution (red dashed line) and PINN inference (blue dashed line). }
    \label{fig:profil1}
\end{figure}

\clearpage

In Fig. (\ref{fig:profil1}), we compare the reference solution and the PINN inference by examining horizontal profiles at two distinct locations. In both cases, the predictions closely match the solution within the main domain. However, it is worth pointing out that at the boundaries of the $x$-domain, subtle discrepancies between the prediction and the reference become apparent for velocity and vorticity fields. This difference can likely be attributed to the absence of explicitly defined boundary conditions for the PINN model.

\begin{figure}[h]
    \centering
    \includegraphics[scale = 0.4]{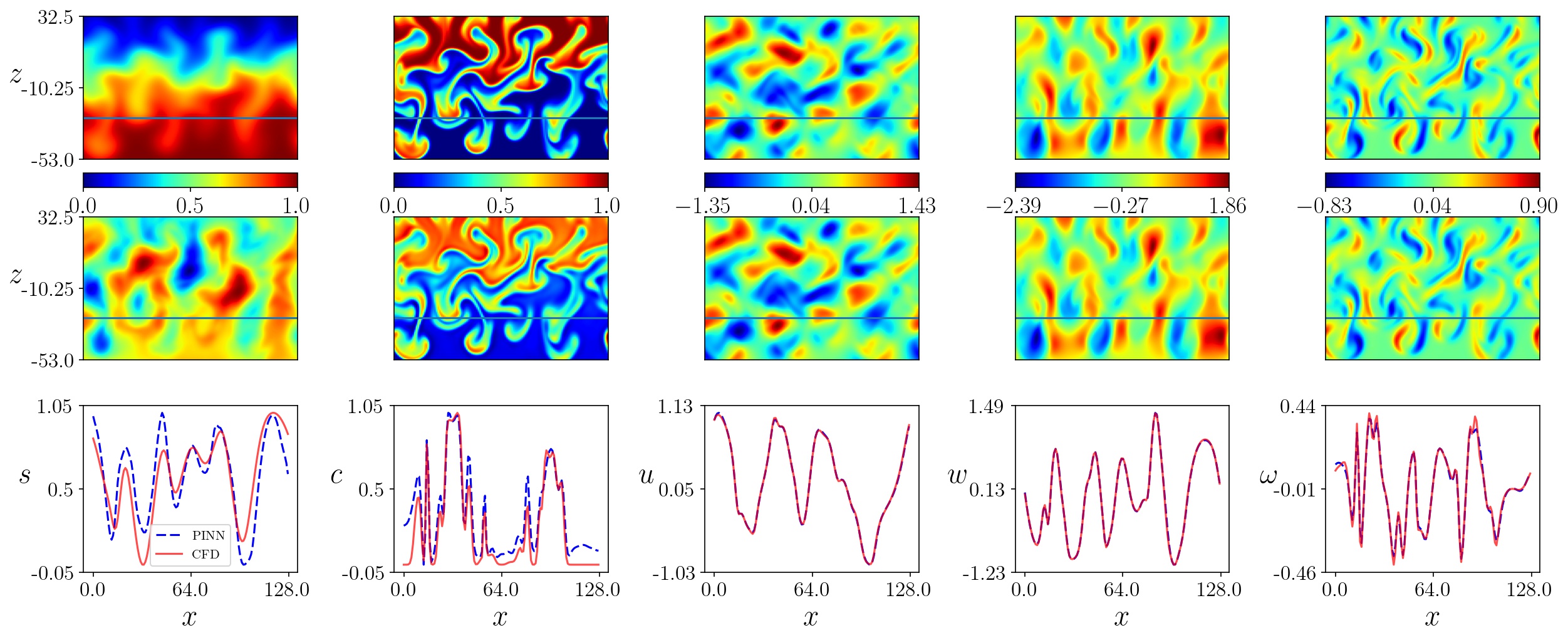}
    \caption{Comparison of horizontal profiles.  Snapshot of salinity, sediment concentration velocity and vorticity at $t = 100$ for the reference solution (first row) and the PINN prediction (seconde row). Horizontal profiles at $z = 28.0$ (red blue line on the snapshots) are compared between CFD reference solution (red dashed line) and PINN inference (blue dashed line). salinity and sediment concentration are normalized using (\ref{eq:minmax}). }
    \label{fig:profil_velocity}
\end{figure}

For the PIV case, the salinity and sediment concentration field predicted cant be compared directly with the corresponding reference field because they are defined to within one constant. Indeed, only the derivatives in time and space of these field are involved in the equations (\ref{eq:sediment_2}). Thus modify the predicted and reference $S$ and $C$ fields as follows :
\begin{align}\label{eq:minmax}
    \hat{V}(\bch{x},t) = \dfrac{V(\bch{x},t) - \inf_{(\bch{x},t)\in X} V}{\sup_{(\bch{x},t)\in X} V- \inf_{(\bch{x},t)\in X} V}
\end{align}

By normalizing the data, it becomes feasible to assess the comparison between these fields, this comparison can be seen Fig. (\ref{fig:profil_velocity}).  While the velocity and the vorticity fields match perfectly with the corresponding references fields, it is hard to recognize the correspondance between CFD and PINN salinity field. The profiles show a big discrepancy for this field and, although the difference is less marked, the sediment concentration field also shows many differences in the horizontal profiles. These results can partially explain the difference in quality between PILAT and PIV case for the identification of $V_p$.

\begin{figure}[h]
    \centering
    \includegraphics [scale = 0.35]{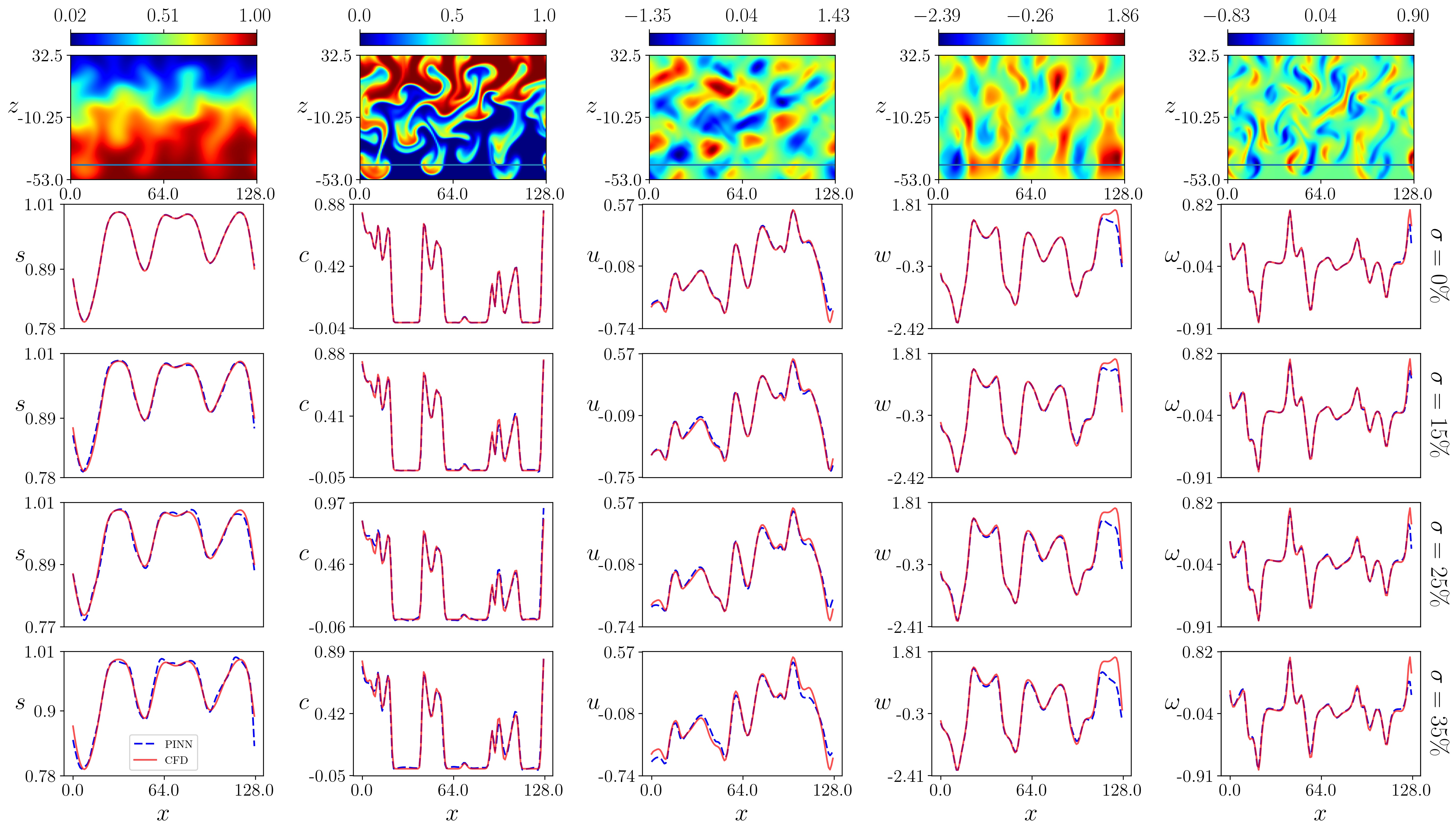}
    \caption{Comparison of horizontal profiles. Snapshots of salinity, sediment concentration velocity and vorticity at $t = 100$ for the reference solution (first row). Horizontal profiles at $z = -43$ (blue line on the snapshots) are compared between CFD reference solution (red line) and PINN inference (blue dashed line) for varying level of gaussian noise where $\sigma = 15\%,~ 25\% $ and $35\%$ from second to last line. }
    \label{fig:noise_profiles}
\end{figure}

For PILAT case, the effect on noise can be seen Fig. (\ref{fig:noise_profiles}), in this figure, we observe the evolution of the same profile as more noise is introduced. The model exhibits a high degree of robustness, however, noticeable discrepancies emerge, particularly around $x$ = 128 for the $w$-component of velocity. Even in the absence of noise, the model struggles to reach the maximum values. Moreover, as the noise level increases, the disparities between the predicted values and the ground truth become more pronounced.

\begin{figure}[h]
     \centering
     \begin{subfigure}[h]{0.49\textwidth}
         \centering
        \includegraphics[width=\textwidth]{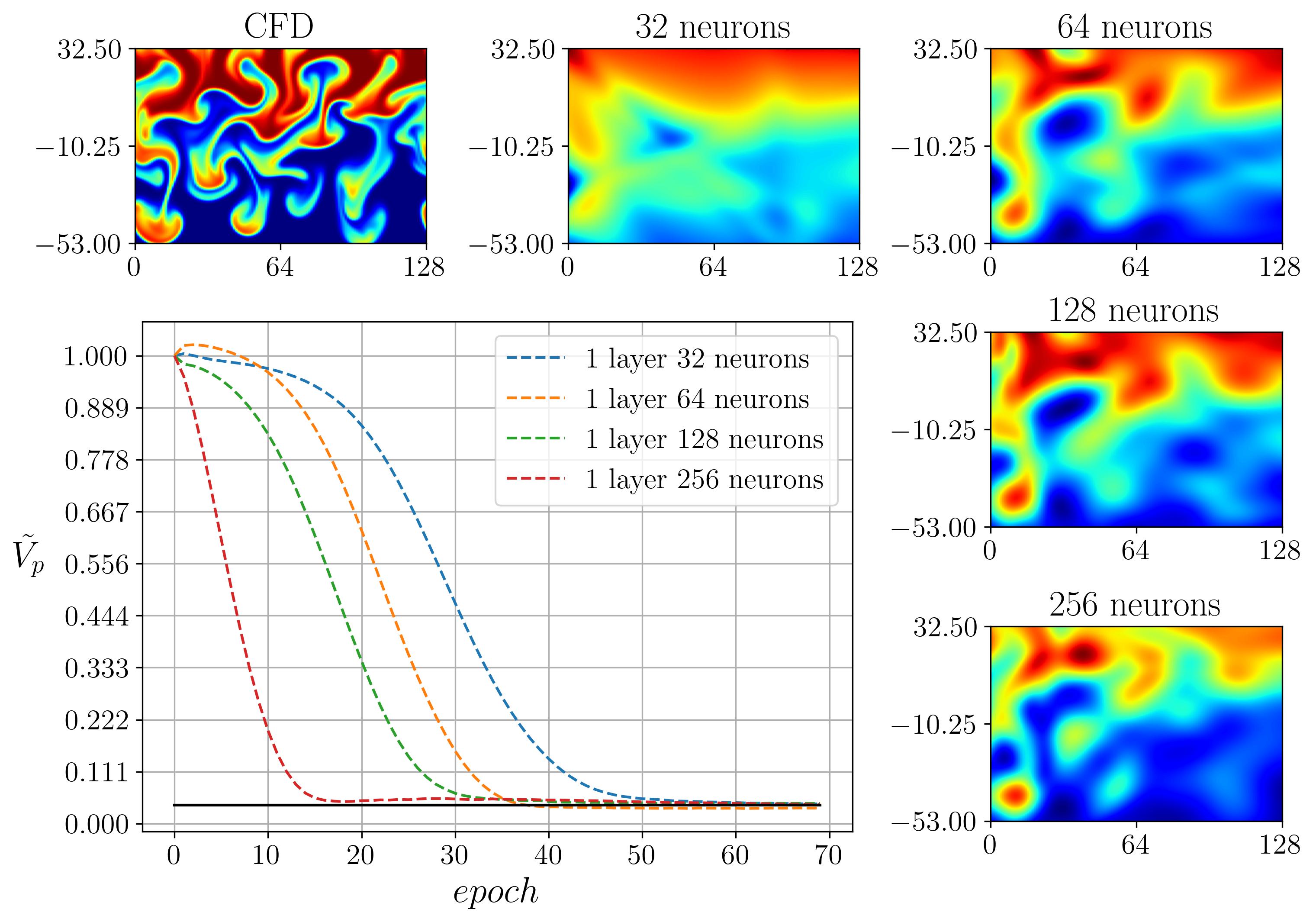}
        \caption{One hidden layer architecture.}
        \label{fig:layer1}
     \end{subfigure}
     \hfill
     \begin{subfigure}[h]{0.49\textwidth}
         \centering
         \includegraphics[width=\textwidth]{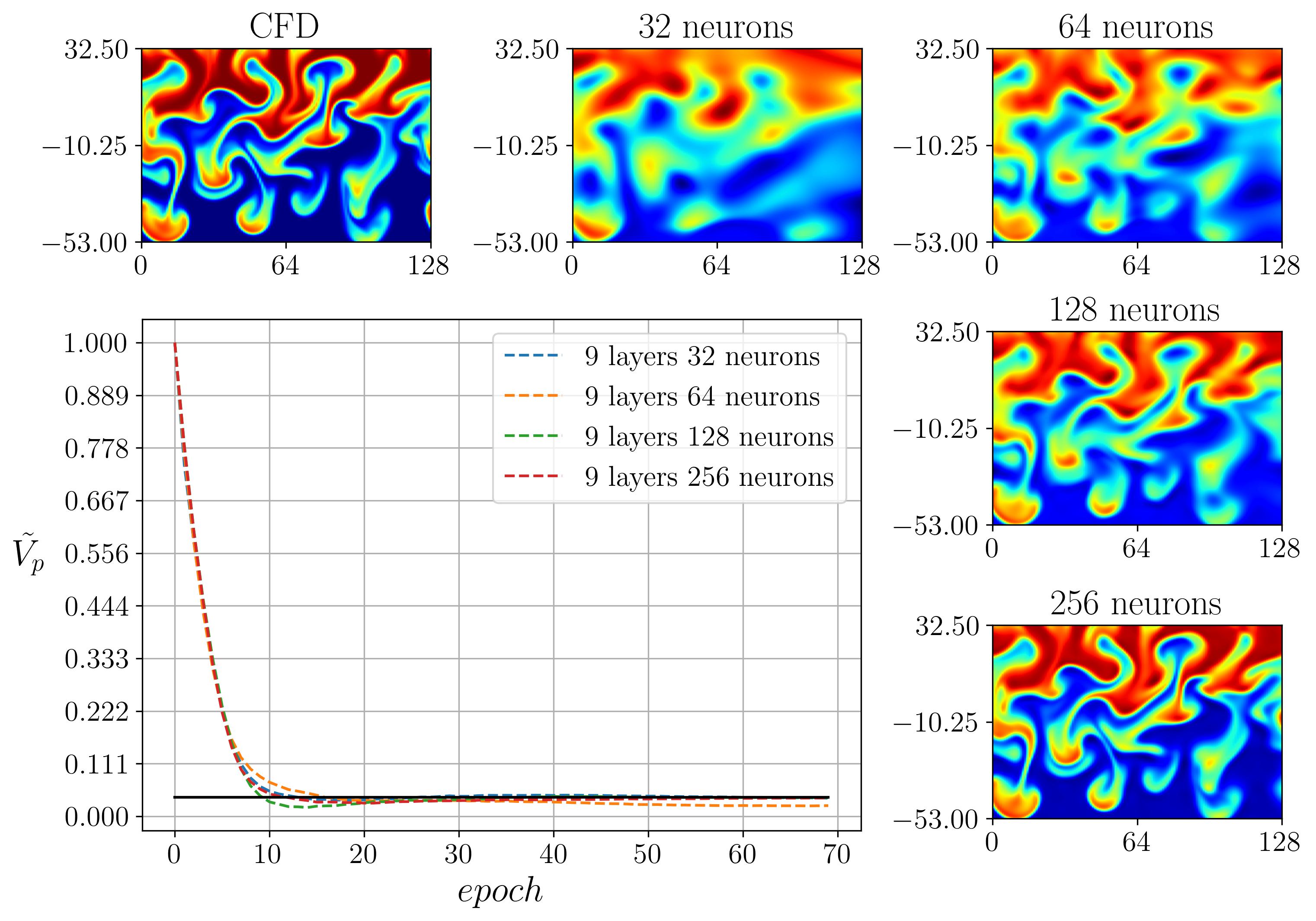}
         \caption{Nine hidden layers architecture.}
         \label{fig:layer9}
     \end{subfigure}
    \caption{Analyzing diverse network structures : architectures with gradually increasing neurons (32, 64, 128, 256 units). For each case, we monitore the evolution of $\Tilde{V_p}$ during training and assessing sediment concentration predictions at t=100 against the reference solution. }
    \label{fig:layers}
\end{figure}

In the following, we study how the quality of $V_p$ identification evolves with networks of varying architectures. In Fig. (\ref{fig:layers}), we can see the evolution of predicted $\Tilde{V_p}$ during training for different number of neurons per layers. We can also qualitatively compare the reconstruction of the different models. In Fig. (\ref{fig:layer1}) a one hidden layer deep network is used while in Fig. (\ref{fig:layer9}), a nine hidden layers deep architecture is investigated. As we can see, all models seems to converge to the reference value $V_p = 0.04$. In the single-layer configuration, the more units, the longer the convergence time. Moreover, for each case the one layer architecture performs poor results on the reconstruction of $C$. In the other hand, For 9 layers, the convergence is slightly faster than for 1 layer, but is the same regardless of the number of neurons. The reconstruction is more accurate than for 1 layers but only 9 layers with 256 units can perform a good reconstruction while inferrinf the right value of $V_p$. 

While each model successfully captures the $V_p$ parameter, the ability to achieve accurate reconstruction holds significant importance. In practical experiments, the capacity to reconstruct effectively serves as a valuable indicator for validating the identification process. Conversely, we would hesitate to place trust in the identification of a model that fails to adequately reconstruct the training fields.

\begin{figure}[h]
    \centering
    \includegraphics[scale = 0.5]{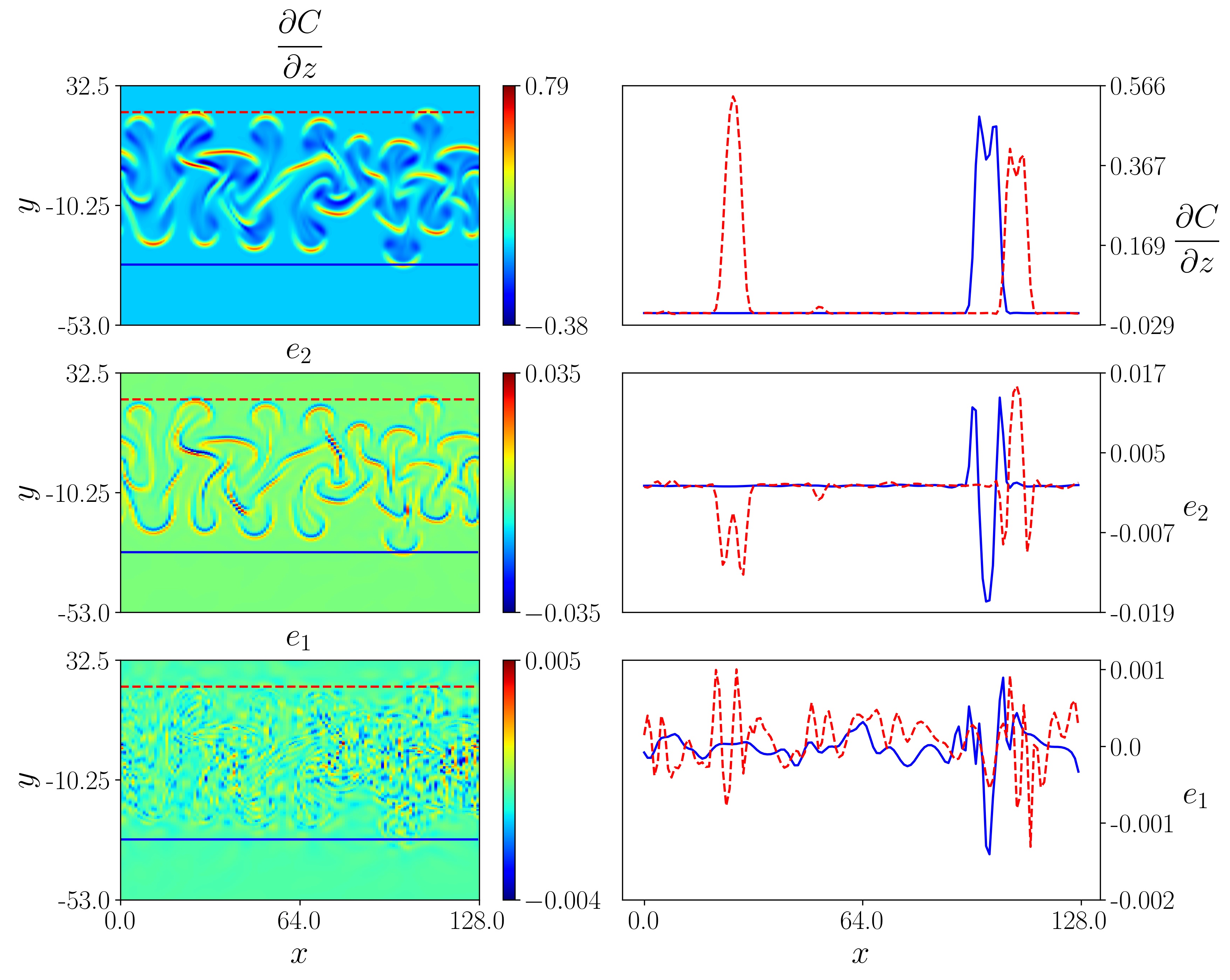}
    \caption{Visualisation of $\dfrac{\partial C}{\partial z}$, and residuals $e_2$ and $e_1$ at $t=100$ (First column). Corresponding horizontal profiles at $z=-28$ and $z=28$ (Second column).}
    \label{fig:partialC}
\end{figure}

Fig. (\ref{fig:partialC}) presents a snapshot at $t=100$ depicting the residuals $e_1$ and $e_2$ in comparison to the partial derivative of $C_{\mt{N}}$ with respect to $z$. In addition, horizontal profiles at two different locations. The residual $e_2$ corresponds to the vorticity equation and exhibits minimal values. Conversely, the residual $e_2$, which is related to sediment concentration transport equation involving $V_p$, displays values ten times higher. Notably, these residuals appear to share a similar shape with $\frac{\partial C_{\mt{N}}}{\partial z}$. Moreover, in the horizontal profile, we can see that pic values for $\frac{\partial C_{\mt{N}}}{\partial z}$ corresponds to pic values for this residuals. As $\Tilde{V_p}$ infered is directly linked to $\frac{\partial C_{\mt{N}}}{\partial z}$. This means that the shape of this residual might be caused by the error of prediction of $\Tilde{V_p}$. In this cases, this is a key result, because the shape of the residuals can be an indirect way to monitor the quality of $V_p$ identification. 

\clearpage

\section*{Conclusion}

This paper demonstrates the effectiveness of Physics-Informed Neural Networks (PINNs) in inferring dimensionless settling velocity in sedimentation flows using synthetic data. Beyond addressing a challenging reconstruction problem involving the coupling of Navier-Stokes equations with the evolution of sediment concentration and salinity, the study successfully identify settling velocity across various flow regimes that illustrate distinct key features of sedimentation processes.
The robustness and applicability of this framework are validated through two sets of training data corrupted with varying levels of noise, one containing salinity and sediment concentration fields and the other containing velocity fields. These results showcase the versatility and reliability of PINNs, particularly in experimental contexts.

\clearpage

\bibliographystyle{elsarticle-num} 
\bibliography{biblio.bib}
\end{document}